\documentclass[aps,prl,twocolumn,superscriptaddress,showpacs,floatfix]{revtex4-1}
\usepackage{graphicx}
\usepackage{color}
\usepackage{bm}
\usepackage{amsmath}
\newcommand{\angstrom}{\text{\normalfont\AA}}
\usepackage{mathrsfs}
\usepackage{multirow}
\usepackage{braket}
\usepackage{comment}
\begin{document}

\title{Conceptual and practical bases for the high accuracy of machine learning interatomic potentials}
\author{Akira \surname{Takahashi}}
\email{takahashi.akira.3n@kyoto-u.ac.jp}
\affiliation{Department of Materials Science and Engineering, Kyoto University, Kyoto 606-8501, Japan}
\author{Atsuto \surname{Seko}}
\email{seko@cms.mtl.kyoto-u.ac.jp}
\affiliation{Department of Materials Science and Engineering, Kyoto University, Kyoto 606-8501, Japan}
\affiliation{Center for Elements Strategy Initiative for Structure Materials (ESISM), Kyoto University, Kyoto 606-8501, Japan}
\affiliation{Center for Materials Research by Information Integration, National Institute for Materials Science, Tsukuba 305-0047, Japan}
\affiliation{JST, PRESTO, Kawaguchi 332-0012, Japan}
\author{Isao \surname{Tanaka}}
\affiliation{Department of Materials Science and Engineering, Kyoto University, Kyoto 606-8501, Japan}
\affiliation{Center for Elements Strategy Initiative for Structure Materials (ESISM), Kyoto University, Kyoto 606-8501, Japan}
\affiliation{Center for Materials Research by Information Integration, National Institute for Materials Science, Tsukuba 305-0047, Japan}
\affiliation{Nanostructures Research Laboratory, Japan Fine Ceramics Center, Nagoya 456-8587, Japan}

\date{\today}
\begin{abstract}
Machine learning interatomic potentials (MLIPs) based on a large dataset obtained by density functional theory (DFT) calculation have been developed recently.
This study gives both conceptual and practical bases for the high accuracy of MLIPs, although MLIPs have been considered to be simply an accurate black-box description of atomic energy.
We also construct the most accurate MLIP of the elemental Ti ever reported using a linearized MLIP framework and many angular-dependent descriptors, which also corresponds to a generalization of the modified embedded atom method (MEAM) potential.
\end{abstract}
\maketitle

\begin{figure}[tbp]
	\includegraphics[clip,width=0.87\linewidth]{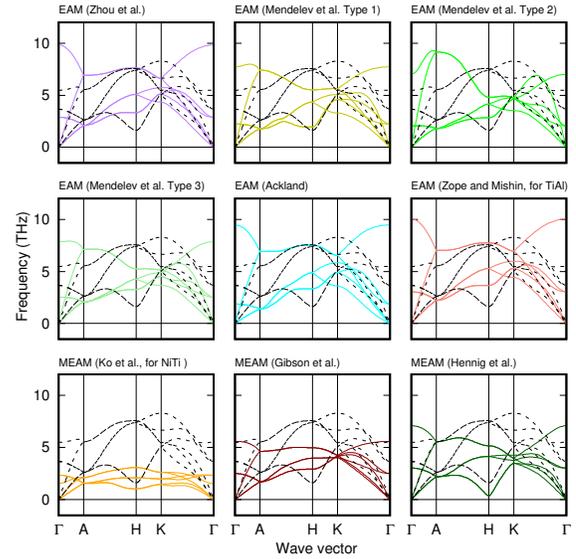}
	\caption{
		Phonon dispersion curves of elemental HCP Ti calculated using conventional EAM\cite{PhysRevB.69.144113,MendelevTitaniumPotential,AcklandManybodyPotential,PhysRevB.68.024102} and MEAM\cite{PhysRevB.92.134107,Zhang2016204,MO_309653492217_000,
PhysRevB.78.054121} potentials.
		Some of these curves are obtained from the interatomic potential repository project\cite{interatomicPotentialRepository} and KIM project\cite{KIMproject}.
		Black broken lines indicate the phonon dispersion curves obtained by DFT calculation.
		Force constants are calculated using the \textsc{lammps}\cite{PLIMPTON19951} code.
	}
		
	\label{_FIG_CONVENTIONAL_PHONON_}
\end{figure}

Interatomic potentials (IPs) have played a central role in performing atomistic simulations, such as molecular dynamics simulation.
A wide variety of conventional IPs have been developed by considering the nature of chemical bonding in specific systems of interest, such as Lennard-Jones\cite{lennard1924determination}, embedded atom method (EAM)\cite{PhysRevLett.50.1285,PhysRevB.29.6443, DAW1993251}, modified EAM (MEAM)\cite{PhysRevLett.59.2666,PhysRevB.46.2727}, and Tersoff\cite{tersoff1986new,PhysRevB.38.9902,PhysRevLett.61.2879} potentials.
However, the accuracy and transferability of conventional IPs are often lacking owing to the simplicity of their potential forms.
As an example, the phonon dispersion relationships of hexagonal close-packed (HCP) Ti computed from several EAM and MEAM potentials are shown in Fig. \ref{_FIG_CONVENTIONAL_PHONON_}, along with that computed on the basis of the density functional theory (DFT).
The overall phonon dispersions of EAM and MEAM potentials are scattered and markedly deviate from that obtained by DFT calculation.

On the other hand, the machine learning IP (MLIP) based on a large dataset obtained by DFT calculation has great potential for improving its accuracy and transferability effectively.
Once the MLIP is established, it does not increase the order of computational cost as compared with conventional IPs.
The MLIP has also been increasingly applied to a wide range of materials regardless of their type of chemical bonding.
Its frameworks applicable to periodic systems have recently been proposed\cite{behler2007generalized,bartok2010gaussian,PhysRevB.90.024101}.

Although the MLIP can provide an accurate energy description, its physical interpretation or relationship with the existing IPs is still lacking.
In this study, we introduce an interpretation of the MLIP on the basis of the framework of EAM and MEAM potentials. 
The interpretation provides a conceptual basis for the high accuracy of the MLIP.
Secondly, we develop the most accurate MLIP of the elemental Ti ever reported using a linearized MLIP framework.
As shown later, the high accuracy of the linearized MLIP implies that the high accuracy and transferability of MLIPs are based mainly on the use of a large number of relevant descriptors, although it has been considered that the use of flexible black-box functions, such as neural network and Gaussian process models, is essential for modeling atomic energy.

The framework of EAM potentials is based on the concept of the embedding energy of an atom into a host described by electron density\cite{PhysRevB.22.1564}.
The embedding energy of atom $i$ is defined as a functional of the host electron density $\rho(r)$ expressed as
\begin{equation}
	E^{(i)}=\mathscr{F}^{(i)}\left[\rho({\bm r})\right],
	\label{_EQ_SCOTT_ZAREMBA_}
\end{equation}
where $\mathscr{F}^{(i)}$ denotes the embedding energy functional for atom $i$.
Although the application of this concept is not exclusive to metallic systems, the framework of EAM potentials is compatible only with metallic systems owing to the introduction of some approximations.
A main approximation is the uniform density approximation (UDA), in which the embedding energy is assumed to be a function of the scalar local electron density, written as
\begin{equation}
	E^{(i)} = F\left(\rho({\bm r}_{i})\right),
	\label{_EQ_UDA_}
\end{equation}
where ${\bm r}_{i}$ denotes the position of atom $i$.
Another one is a pairwise approximation in which the local electron density is assumed to be equal to the sum of contributions from neighboring atoms expressed by a single pairwise function.
Adding a short-range pairwise interaction, the EAM atomic energy is expressed as
\begin{equation}
	E^{(i)} = F\left(\sum_j p(r_{ij}) \right) + \frac{1}{2}\sum_j \phi(r_{ij}),
	\label{_EQ_EAM_}
\end{equation}
where $p(r_{ij})$ and $\phi(r_{ij})$ denote the pairwise contribution of the neighbor atom $j$ to the local electron density and short-range pairwise interaction including repulsive energy, respectively\cite{PhysRevLett.50.1285}.
In an extended manner, the MEAM atomic energy is given by
\begin{equation}
	E^{(i)} = F\left(\rho(\bm{r}_i)\right) + \frac{1}{2}\sum_j \phi(r_{ij}),
	\label{_EQ_MEAM_}
\end{equation}
\begin{equation}
	\rho(\bm{r}_i)=\sum_j p(r_{ij}) + \sum_{j,k} f(r_{ij})f(r_{ik})g(\cos\gamma_{jik}),
	\label{_EQ_MEAM2_}
\end{equation}
where the local electron density is described by a three-body function $g$ in addition to the pairwise contribution.
Since the function forms of $p$, $f$, and $g$ have not been established, a wide range of approximated forms have been proposed in the literature.
In addition, polynomials and spline models have been simply used as function $F$.

On the other hand, all MLIPs with pairwise descriptors are formulated as
\begin{equation}
	E^{(i)}=F\left(b_{10}^{(i)},b_{20}^{(i)},\dots,b_{n_{\rm{max}0}}^{(i)}\right),
\label{_EQ_MLIP_RAD_}
\end{equation}
where $b_{n0}^{(i)}$ denotes a pairwise descriptor expressed as
\begin{equation}
	b_{n0}^{(i)}=\sum_j f_n(r_{ij}).
	\label{_EQ_PAIR_DESCRIPTOR_}
\end{equation}
A large number of pairwise descriptors are generally used for formulating MLIPs, and neural network models, Gaussian process models, and polynomials have been used as functions $F$. 
This formulation is obviously a generalization of the EAM atomic energy.
Similarly, most MLIPs with angular-dependent descriptors are formulated as
\begin{equation}
	E^{(i)} = F(b_{10}^{(i)}, b_{20}^{(i)}, \dots, b_{11}^{(i)}, b_{21}^{(i)}, \dots, b_{n_{\rm max}l_{\rm max}}^{(i)}),
	\label{_EQ_MLIP_ANGLE_}
\end{equation}
where $b_{nl}^{(i)}$ denotes an angular-dependent descriptor.
Most angular-dependent descriptors specified by number $l$ belong to the class of angular Fourier series, which corresponds to a set of rotationally invariant descriptors derived from spherical harmonics\cite{bartok2013representing}.
The angular Fourier series is given by
\begin{equation}
	b_{nl}^{(i)}=\sum_{j,k}f_n(r_{ij})f_n(r_{ik})\cos^l\left(\gamma_{jik}\right) (l \geq 1),
	\label{_EQ_DESCRIPTORS_ANGLE_}
\end{equation}
where $\gamma_{jik}$ denotes the bond angle between atoms $j-i-k$.
From the comparison between Eqns.(\ref{_EQ_MEAM_}) and (\ref{_EQ_MLIP_ANGLE_}), the formulation of the MLIP with angular-dependent descriptors is clearly a generalization of the MEAM potential.

We have demonstrated that the MLIP formulations can be regarded as the generalizations of the EAM and MEAM potentials by comparing their equations for atomic energy.
We will show that the MLIP formulations can also be derived from the concept of embedding energy using a higher-order approximation beyond the UDA.
This derivation interprets MLIPs.
Using a higher-order approximation for the embedding energy functional (Eqn.(\ref{_EQ_SCOTT_ZAREMBA_})), atomic energy may be described by a function of local electron density and its derivatives as
\begin{equation}
\begin{split}
	E^{(i)}= & \mathscr{F}^{(i)}\left[\rho({\bm r})\right] \\
	=&F\left(\rho(\bm{r}_i), 
	\frac{\partial\rho}{\partial x}({\bm r}_i),
	\frac{\partial\rho}{\partial y}({\bm r}_i),
	\frac{\partial\rho}{\partial z}({\bm r}_i),
	\dots
	\right).
\end{split}
\label{_EQ_EAM_GRADIENT_}
\end{equation}
Then, the local electron density is assumed to be described by direction-dependent contributions from neighbor atoms, $\rho(\bm{r}_i)=\sum_jp(\bm{r}_{ij})$.
Eqn.(\ref{_EQ_EAM_GRADIENT_}) is rewritten as
\begin{widetext}
\begin{equation}
	E^{(i)}=
	F\left(\sum_jp(\bm{r}_{ij}), 
	\sum_j\frac{\partial}{\partial x}p({\bm r}_{ij}),
	\sum_j\frac{\partial}{\partial y}p({\bm r}_{ij}),
	\sum_j\frac{\partial}{\partial z}p({\bm r}_{ij}),
	\dots
	\right).
\label{_EQ_EAM_GRADIENT2_}
\end{equation}
\end{widetext}
Expanding the electron density contribution $p$ using a basis set $\{f_n(\bm{r}_{ij})\}_{n=1,2,\dots,n_{\rm max}}$ as
\begin{equation}
	p(\bm{r}_{ij})=\sum_{n=1}^{n_{\rm max}}c_nf_n(\bm{r}_{ij}),
	\label{_EQ_ELECTRON_EXPAND_}
\end{equation}
embedding atomic energy is written as 
\begin{equation}
	E^{(i)}=
	\tilde{F}\left(\sum_j f_1(\bm{r}_{ij}), 
	\dots,
	\sum_j f_{n_{\rm max}}(\bm{r}_{ij})
	\right),
\label{_EQ_MLIP_}
\end{equation}
where another symbol $\tilde{F}$ for the embedding energy function is derived from both function $F$ and expansion coefficients $\{c_n\}_{n=1,2,\dots,n_{\rm max}}$.
Replacing the vector $\bm{r}_{ij}$ with the pair distance $r_{ij}$, Eqn.(\ref{_EQ_MLIP_}) becomes the pairwise MLIP formulation.
Generally, the basis set is not necessarily pairwise.
When functions based on spherical harmonics are used as a basis set and function $\tilde{F}$ satisfying the rotational invariance, the angular-dependent MLIP (Eqn.(\ref{_EQ_MLIP_ANGLE_})) is derived.
Thus, MLIP formulations are derived from the concept of embedding energy using an approximation beyond the UDA.
This implies that the lack of accuracy and transferability of the EAM and MEAM potentials can be ascribed to their poor representation for embedding energy due to the limitation of the UDA\footnote[1]{
	Even if $p(r_{ij})$ is expressed by the linear combination of two functions $f_1(r_{ij})$ and $f_2(r_{ij})$, Eqn.(\ref{_EQ_EAM_}) is not enough to express some functions such as $\left[\sum_j f_1(r_{ij})\right]^2 + \left[\sum_j f_2(r_{ij})\right]^2$.}.

On the basis of the relationship between MLIPs and EAM potentials, we construct two MLIPs for the elemental Ti in this study. 
The first one is constructed by a third-order polynomial approximation of Eqn.(\ref{_EQ_MLIP_RAD_}) expressed as 
\begin{eqnarray}
	\nonumber	E^{(i)} = w_0 + \sum_n\left[ w_{n} b^{(i)}_{n0}\right] + \sum_{n,n'}\left[ w_{n,n'} b^{(i)}_{n0} b^{(i)}_{n'0}\right] + \\
	\sum_{n,n',n''} \left[w_{n,n',n''} b^{(i)}_{n0} b^{(i)}_{n'0} b^{(i)}_{n''0}\right],
	\label{_EQ_POT_RADIAL_}
\end{eqnarray}
where $w_0, w_n, w_{n,n'},$ and $w_{n,n',n''}$ denote regression coefficients.
The second one is constructed by a second-order polynomial approximation of Eqn.(\ref{_EQ_MLIP_ANGLE_}) with angular Fourier series descriptors expressed as 
\begin{equation}
	E^{(i)} = w_0 + \sum_{n,l} \left[ w_{n,l} b^{(i)}_{nl} \right] + \sum_{n,l,n',l'} \left[ w_{n,l,n',l'} b^{(i)}_{nl} b^{(i)}_{n'l'} \right].
	\label{_EQ_POT_ANGULAR_}
\end{equation}
Here, we fixed $l_{\rm max}$ to ten.
We used pairwise Gaussian-type functions as radial functions $f_n(r)$ expressed as
\begin{equation}    
f_{n}(r)=f_{c}(r)\exp\left[-p(r-q_{n})^{2}\right],
\end{equation}
where $f_c (r)$ denotes a cosine-type cutoff function.
$p$ and $q_n$ are given parameters, and we used a single $p$ value and a set of $q_n$ values given by an arithmetic sequence.
Also in the EAM and MEAM potentials, Gaussian functions have sometimes been used for expressing the pairwise electron density contribution. 
In addition, a polynomial approximation for the embedding energy function $F$ has been used for EAM and MEAM potentials.
Therefore, the only difference between the MLIP and EAM (MEAM) potentials is in the number of descriptors being used in the formulation of atomic energy.
Eqns.(\ref{_EQ_POT_RADIAL_}) and (\ref{_EQ_POT_ANGULAR_}) are also a generalization of our previous linearized model where only the power of $b_n$ is considered\cite{PhysRevB.90.024101,PhysRevB.92.054113}.

Training and test datasets were generated by DFT calculation for 2700 and 300 atomic configurations, respectively.
We firstly optimized the atomic positions and lattice constants of face-centered cubic (FCC), body-centered cubic (BCC), HCP, simple cubic (SC), $\omega$, and $\beta$-Sn structures, and supercells were then developed by the $2\times2\times2$, $3\times3\times3$, $3\times3\times3$, $4\times4\times4$, $3\times3\times3$, and $2\times2\times2$ expansions of their conventional unit cells, respectively.
Atomic configurations were generated by isotropic expansion, random expansions, random distortions, and random displacements. 
Both the energy and forces acting on each atom were used for training.
Therefore, the total number of training data was 430650. 

We adopted linear ridge regression to estimate MLIPs involving the minimization of a function defined by the energy and forces acting on atoms.
The function is defined elsewhere\cite{PhysRevB.92.054113}.
DFT calculation was performed using the plane-wave basis projector augmented wave (PAW) method\cite{PAW1,PAW2} within the Perdew--Burke--Ernzerhof exchange-correlation functional\cite{GGA:PBE96} as implemented in the \textsc{vasp} code\cite{VASP1,VASP2}.
The cutoff energy was set to 400 eV. 
The total energies converged to less than $10^{-3}$ meV/supercell.
The lattice constants of the ideal structures were optimized until the residual forces became less than $10^{-3}$ eV/\AA.

\begin{figure}[tbp]
	\includegraphics[clip,width=0.95\linewidth]{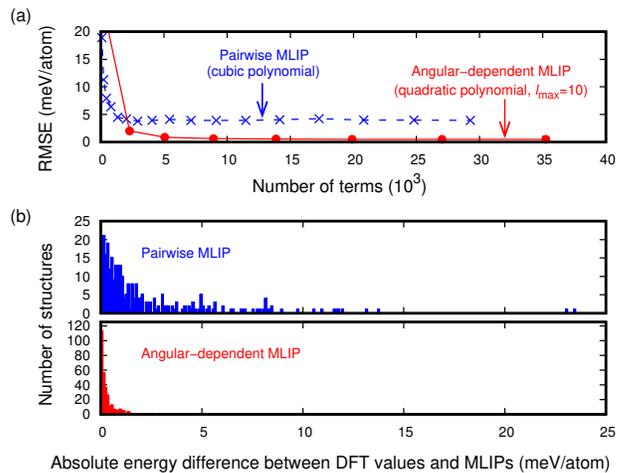}
	\caption{
		(a) Dependence of RMSE of MLIPs on number of terms for elemental Ti.
		(b) Distribution of absolute energy difference between DFT values and MLIPs.
	}
	\label{_FIG_RMSE_}
\end{figure}

We will show the accuracy of MLIPs for the elemental Ti.
We regard the root mean square error (RMSE) for the energy of the test dataset as a measure of prediction error.
Figure \ref{_FIG_RMSE_} (a) shows the dependence of prediction error on the number of regression coefficients.
The number of regression coefficients was controlled using only the number of radial functions $f_n$ for both pairwise and angular-dependent MLIPs. By examining the convergence of RMSE with respect to the number of regression coefficients, we obtained an optimized pairwise MLIP with a prediction error of 3.8 meV/atom (2925 coefficients).
Similarly, we obtained an optimized angular-dependent MLIP with a prediction error of 0.5 meV/atom (35245 coefficients), which means that it is very important to consider angular-dependent descriptors for expressing the interatomic interactions of the elemental Ti.
Figure \ref{_FIG_RMSE_} (b) also shows the distribution of the absolute energy difference between DFT and MLIPs for the test dataset.
The distribution for the angular-dependent MLIP is much narrower than that for the pairwise MLIP, which is consistent with the degree of prediction error.
For the angular-dependent MLIP, more than a hundred structures show the absolute energy difference within only 0.1 meV/atom.
In addition, some outliers can be found in the distribution for the pairwise MLIP.
A structure shows the maximum absolute energy difference of 23.0 meV/atom of the pairwise MLIP, whereas the absolute energy difference of the angular-dependent MLIP does not exceed 2.8 meV/atom.

\begin{figure}[tbp]
	\includegraphics[clip,width=0.95\linewidth]{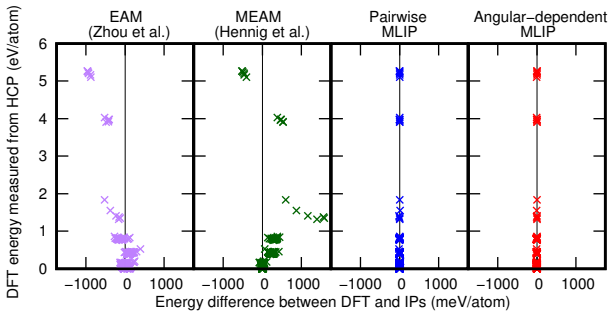}
	\caption{Distribution of energy difference between DFT and IPs.}
	\label{_FIG_PREDICTION_ERROR_}
\end{figure}

We then compare the distribution of the energy difference between DFT and IPs for the test data, elastic constants and phonon dispersion relationships obtained from EAM\cite{PhysRevB.69.144113} and MEAM\cite{PhysRevB.78.054121} potentials, the pairwise MLIP and the angular-dependent MLIP along with a reference of the DFT calculation.
Figure \ref{_FIG_PREDICTION_ERROR_} shows the comparison of the distribution of energy difference between DFT and IPs for the test  dataset.
EAM and MEAM potentials show very large energy differences for almost the entire test dataset, while both the MLIPs show very small energy differences.

\begin{figure}[tbp]
	\includegraphics[clip,width=0.95\linewidth]{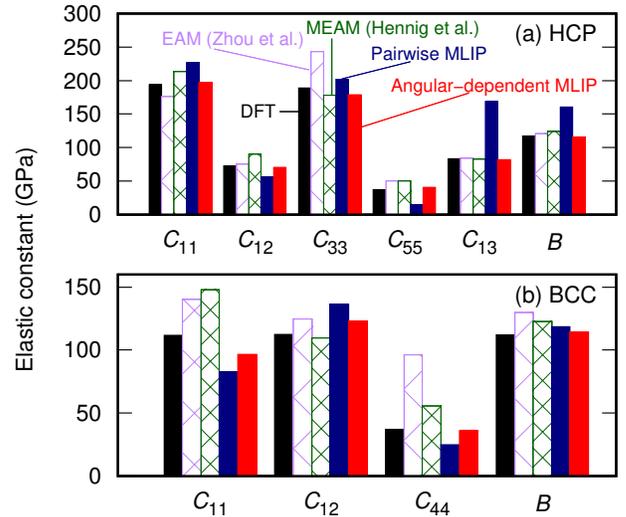}
	\caption{Elastic constant $C$ and bulk modulus $B$ values of (a) HCP-Ti and (b) BCC-Ti calculated on the basis of DFT and IPs.}
	\label{_FIG_ELAS_}
\end{figure}

Figure \ref{_FIG_ELAS_} shows the elastic constants and bulk moduli of (a) HCP-Ti and (b) BCC-Ti obtained from EAM and MEAM potentials and the MLIPs. 
The elastic constants of EAM and MEAM potentials are close to those of DFT calculation, except for the $C_{33}$ of HCP and the $C_{44}$ of BCC obtained from the EAM potential.
On the other hand, the pairwise MLIP is worst for predicting most of the elastic constants and bulk moduli of both HCP and BCC structures, despite its small prediction error.
Including angular-dependent terms, the prediction of elastic constants and bulk moduli is much improved.
This is consistent with the fact that the angular-dependent descriptors are essential for predicting the mechanical behavior of the elemental Ti.

\begin{figure*}[t]
	\includegraphics[clip,width=0.95\linewidth]{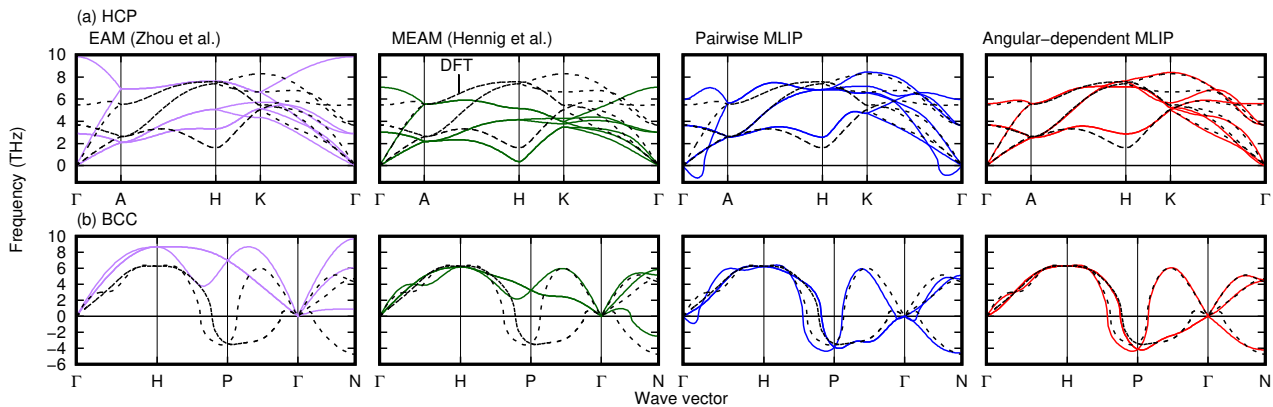}
	\caption{
		Phonon dispersion curves of (a) HCP-Ti and (b) BCC-Ti calculated from IPs.
		Broken black lines indicate the phonon dispersion curves calculated on the basis of DFT.
	}
	\label{_FIG_PHONON_}
\end{figure*}

The phonon dispersion curves were also calculated using the supercell approach\cite{Parlinski} for HCP and BCC structures with the  DFT equilibrium lattice constant.
To evaluate a dynamical matrix, each symmetrically independent atomic position was displaced by 0.01 $\angstrom$.
The forces acting on atoms were then computed.
Supercells were fabricated by the 4 $\times$ 4 $\times$ 4 expansion of conventional unit cells for both HCP and BCC structures.
Phonon calculations were performed using the \textsc{phonopy} code\cite{Togo20151}.
Figure \ref{_FIG_PHONON_} shows the phonon dispersion curves of (a) HCP and (b) BCC structures computed from EAM and MEAM potentials, and the MLIPs.
As shown in Fig. \ref{_FIG_PHONON_}, the phonon dispersion curves from EAM and MEAM potentials differ largely from that obtained by DFT calculation.
Imaginary phonon modes are observed in the DFT phonon dispersion for the BCC structure, but not in the EAM and MEAM phonon dispersions.
Although the pairwise MLIP reproduces the DFT phonon dispersion better than the EAM and MEAM potentials, phonon frequencies tend to be overestimated.
The angular-dependent MLIP significantly improves the inconsistency of phonon frequency.

In summary, this study provides both conceptual and practical bases for the high accuracy of MLIPs.
We have shown that MLIPs can be regarded as a description of embedding energy beyond the UDA, which is a fundamental approximation of both EAM and MEAM potentials.
In other words, the high accuracy of MLIPs is based on the use of higher-order approximation of embedding energy.
We have then applied a linearized MLIP approach to the elemental Ti, which is also a generalization of the MEAM potential.
An angular-dependent linearized MLIP predicts the energetics and phonon frequencies much more accurately than the existing MEAM potentials.
The only difference between the MEAM potentials and linearized MLIP is in the number of descriptors being used.
This indicates that the use of a systematic set of numerous descriptors is the most important practical feature for building MLIPs with high accuracy.

\section{Acknowledgements}

This study was supported by PRESTO from JST, a Grant-in-Aid for Scientific Research (B) (Grant No. 15H04116) from JSPS and a Grant-in-Aid for Scientific Research on Innovative Areas ``Nano Informatics'' (Grant No. 25106005) from JSPS. 
AT was supported by a Grant-in-Aid for JSPS Research Fellows (Grant No. 15J07315) from JSPS. 
\bibliography{takahashi_pot}
\end{document}